\documentclass{PoS}

\usepackage{graphicx}
\usepackage{aas_macros}
\newcommand{\pks}{PKS\,2155-304}

\title{Unraveling The Complex Nature Of The Very High-Energy $\gamma$-Ray Blazar PKS\,2155-304}

\ShortTitle{Nature of PKS\,2155-304}

\author{\speaker{Alicja Wierzcholska}$^{a, b}$, Michael Zacharias$^{c,d}$, Felix Jankowsky$^{a}$, Stefan Wagner$^{a}$ \\
        \llap{$^a$}Landessternwarte, Universit\"at Heidelberg, K\"onigstuhl, D 69117 Heidelberg, Germany \\
        \llap{$^b$}Institute of Nuclear Physics PAN,  ul. Radzikowskiego 152, 31-342 Krakow, Poland\\
        \llap{$^c$}Ruhr Astroparticle and Plasma Physics Center (RAPP Center), Institut fur theoretische Physik IV, Ruhr-Universit\" Bochum, D-44780, Bochum, Germany  \\
        \llap{$^d$}Centre for Space Research, North-West University, Potchefstroom 2520, South Africa\\
        E-mail: \email{alicja.wierzcholska@ifj.edu.pl}}

\abstract{

\pks\ is a blazar located in the Southern Hemisphere, monitored with the High Energy Stereoscopic System (H.E.S.S.) at very high energy (VHE, $E>100$\,GeV) $\gamma$ rays every year since 2002. Thanks to the large data set collected in the VHE range and simultaneous coverage in optical, ultraviolet (UV), X-ray and high energy gamma-ray ranges, this object is an excellent laboratory to study spectral and temporal variability in blazars. However, despite many years of dense monitoring, the nature of the variability observed in PKS 2155-304 remains puzzling.
In this paper, we discuss the complex spectral and temporal variability observed in PKS 2155-304.
The data discussed include VHE $\gamma$-ray data collected with H.E.S.S. between 2013 and 2016, complemented with multiwavelength (MWL) observations from Fermi-LAT, Swift-XRT, Swift-UVOT, SMARTS,  and the ATOM telescope.
During the period of monitoring, \pks\  was transitioning from its lower state to the flaring states, and exhibiting different flavors of outbursts. 
For the first time, orphan optical flare lasting a few months was observed.
Correlation studies show an indication of correlation between the X-ray and VHE $\gamma$-ray fluxes. 
Interestingly, a  comparison of optical and X-ray or VHE $\gamma$-ray fluxes does not show global correlation. 
However, two distinct tracks in the diagram were found, which correspond to the different flaring activity states of \pks.

}

\FullConference{High Energy Phenomena in Relativistic Outflows VII - HEPRO VII\\
		9-12 July 2019\\
		Facultat de Fisica, Universitat de Barcelona, Spain}

\begin{document}

\section{Introduction}
PKS\,2155-304  (redshift z=0.117) is an HBL (high energy-peaked BL Lacertae) type blazar, one of the brightest blazars in  the  Southern  Hemisphere. 
The first observation reporting \pks\ were performed in the radio frequencies as a part of the Parkes survey \cite{Shimmins}.
Since its discovery, \pks\  is a target for observations at different wavelengths for already 40 years  \cite{Zhang2006,  Kastendieck2011, Wierzcholska_swift}.
H.E.S.S. (High Energy Stereoscopic System) started observing \pks\ in 2002 \cite{2155_2005} and since this time as this is a very bright BL Lac object, observations are performed every year. 
From 2004 to 2012, yearly  H.E.S.S. monitoring of the blazar was performed with four Cherenkov telescopes, while in November 2012, the H.E.S.S. experiment was upgraded with the fifth telescope, and since this time monitoring was performed with the array of five. 
These long-term observations unraveled the very complex nature of \pks.
In 2006 an exceptional flare,  with the flux of 40 times the average flux and minutes scale variability was detected \cite{2155_1flare}.
Furthermore, 44 hours later, another flare, with the flux 20$\%$ larger than the first outburst, was observed  \cite{2155_2flare}.
The MWL data collected simultaneously during this second flaring activity revealed a strong X-ray  - $\gamma$-ray flux cubic relation at high flux levels, weakening at lower flux levels.
Interestingly, no optical   $\gamma$-ray flux correlation has been found for the period of the observations. 
In August 2008, H.E.S.S., RXTE, Swift, and ATOM observed \pks\ in its low state simultaneously during 11 nights.
The campaign for the first time revealed an indication for a positive correlation between optical and VHE  $\gamma$-ray fluxes, with the Pearson correlation coefficient of 0.8  \cite{2155_optgam}.
However, the analysis of the VHE data collected from 2007 to 2009 did not show any universal relation between the very high-energy $\gamma$-ray and optical flux changes on the timescales from days and weeks up to several years \cite{2155_atom}.
But, the authors pointed out that at higher flux levels the source can follow two distinct tracks in the optical flux-color diagrams, which seem to be related to distinct $\gamma$-ray states of the blazar.
Further studies on the monitoring data collected between 2004 and 2012  with H.E.S.S. and between 2008 and 2012 with Fermi-LAT have shown that emission observed in the quiescent state of \pks\ is consistent with a log-normal spectral behavior \cite{2155_lognorm}.

\section{Data analysis}
\subsection{Very high energy $\gamma$-ray observations with H.E.S.S.}
H.E.S.S. is an array of five Imaging Atmospheric Cherenkov Telescopes (IACTs), located in the Khomas Highland in Namibia.
The observations were performed in the regime of  very-high-energy  $\gamma$ rays  \cite{Aharonian2006_crab}.
Since 2012, the system consists of four 12\,m telescopes, each with a mirror area of  108\,m$^2$,  a fifth, larger telescope with a mirror area of  614\,m$^2$. 
The H.E.S.S. data analysis was performed using the Model analysis chain \cite{Naurois09} with the Standard Cuts configuration \cite{Aharonian2006_crab}.
Similar results as those presented in the paper were obtained using the independent ImPACT analysis chain \cite{Parsons14}
\subsection{High energy $\gamma$-ray observations with Fermi-LAT}
High energy $\gamma$-ray data (HE,  E$>$100\,MeV) collected with the \emph{Fermi}-LAT in the period of January 1, 2013, and December 31, 2016, have been analyzed using standard Fermi Science Tools (version v10r0p5) with P8R2$\_$SOURCE$\_$V6 instrument response functions   \cite{Atwood13}.
The analysis was performed using all events in the energy range from 100 MeV to 500 GeV.
The region of interest was constrained to have a size of 10\,degree radius and be centered on the source. 
As an analysis method binned maximum-likelihood was applied with the Galactic diffuse background modeled using the \verb|gll_iem_v06| map cube, and the extragalactic diffuse and residual instrument backgrounds modeled jointly using the \verb|iso_P8R2_SOURCE_V6_v06| template \cite{Mattox96}.
\subsection{Monitoring with Swift}
The Neil Gehrels Swift Observatory monitored \pks\ monitors \pks\ in optical, UV and X-ray energy ranges. 
All data collected with Swift-XRT and Swift-UVOT,  with ObsID  00049686001-00049686005, which corresponds to all observations taken in 2013-2016,  were analysed. 
For the X-ray observations, the analysis was performed using \verb|HEASOFT| v.6.23 and for spectral fitting  \verb|XSPEC| v.12.9.1 was used. 
All data were binned in order to have at least 20 counts per bin and fitted using a single power-law model with a Galactic absorption value of 1.52$\cdot$10$^{20}$\,cm$^{-2}$ \cite{Kalberla2005}  set as a fixed parameter. 

In the case of optical/UV observations,  instrumental magnitudes were calculated using the \verb|uvotsource| task.
A region of interest was defined as a circle with a radius of 5\,arcsec. 
The same size region, located close to the source region, but being uncontaminated with any signal from the nearby sources,  was used to estimate background emission.
The flux conversion factors were taken from \cite{Poole08}.
The UVOT data were corrected for the influence of the Galactic extinction based on the model from \cite{Schlegel98} with the most recent recalibration by \cite{Schlafly}, 
using the Extinction Calculator from NED (NASA/IPAC EXTRAGALACTIC DATABASE). 

\subsection{Optical observations with ATOM and SMARTS}
The optical monitoring of \pks\ was also performed with SMARTS (Small and Moderate Aperture Research Telescope System, \cite{Bonning2012}) and ATOM (Automatic Telescope for Optical  Monitoring, \cite{ATOM}).
For the periods 2013-2014 and 2015-2016 SMARTS and ATOM data were used, respectively.

The magnitudes have been corrected for the influence of the Galactic extinction. 
For the spectral studies, all optical data were corrected for the contribution of the host galaxy, using the template of an elliptical galaxy provided by \cite{Fukugita95} and observations in Gunn filter $i$  \cite{Falomo91}, with an assumed de Vaucouleurs profile of the starlight.

\section{Longterm light curve}
\begin{figure*}
 \centering{\includegraphics[width=0.6\textwidth]{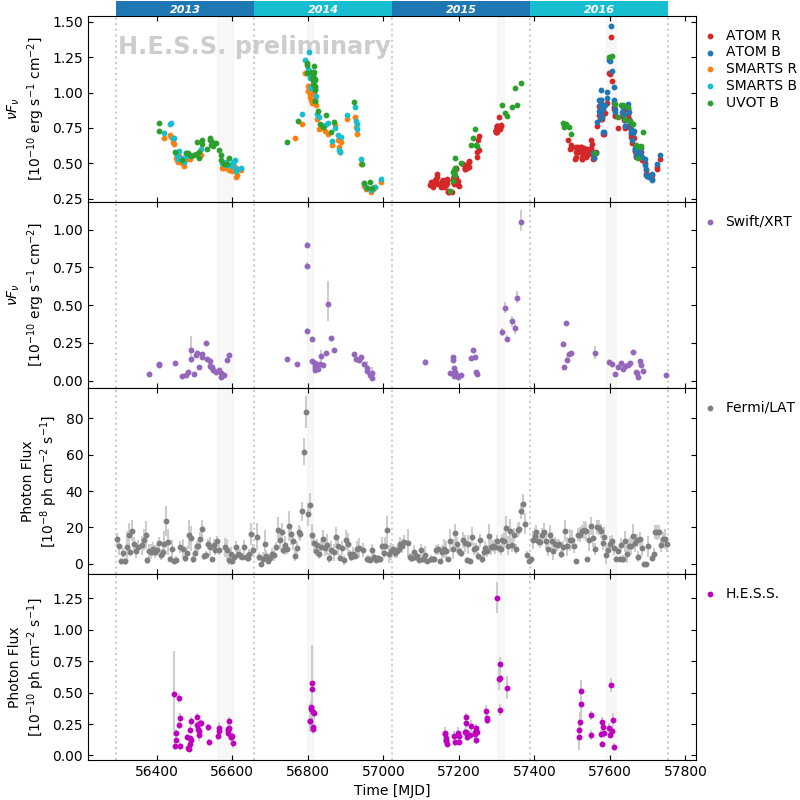}}
 \caption[]{MWL light curve of \pks\ presenting data collected in 2013-2016 with ATOM, SMARTS, Swift-UVOT, Swift-XRT, Fermi-LAT and H.E.S.S.
 The vertical gray areas indicate four intervals selected for the modelling. }
\label{lc}
\end{figure*}

Figure\,\ref{lc} presents the long-term light curve of \pks\ as seen with six different instruments: H.E.S.S., Fermi-LAT, Swift-XRT, Swift-UVOT, SMARTS, and ATOM.
All of the data collected span four years of monitoring of blazar, starting from the beginning of 2013. 
During this period \pks\ was observed in different activity states with variability pattern changing from epoch to epoch. 
Significant variability is visible for all wavebands. 
In particular, in 2013 the blazar was rather quiet and not manifesting large flaring activity. 
But, in May 2014 a huge outburst was observed in all frequencies.  Unfortunately, H.E.S.S. monitoring started only in the decay phase of the outburst. 
The second flare observed in 2014 was visible only in optical and UV frequencies. 
There was no simultaneous activity observed either in the  X-ray band or HE $\gamma$-ray one. Unfortunately, there are no VHE $\gamma$-ray observations for this period. 
In 2015 \pks\ was also active, with a prominent outburst observed at all frequencies. 
In 2016 \pks\ was rather quiet in all energy bands except for optical and UV ones. 
In these regimes, large, lasting few months outburst was detected.

\section{Correlation studies}
The dataset presented in Figure \ref{lc} provides a perfect possibility to study MWL correlations. 
Comparison of the optical B-R color vs. B band fluxes do not show any trend in the diagram. 
Correlation between optical color and flux has previously reported for the optical observation of \pks\ collected in the period 2007-2012 \cite{2155_optgam, 2155_atom}.
However, the authors pointed out that a clear bluer-when-brighter trend is not visible in the entire set of data studied, but only in sub-branches. 

Indication of the correlation has been found in the comparison of X-ray and VHE $\gamma$-ray data. Pearson correlation coefficient for this matching is equal to 0.76$\pm$0.19.
The presence of X-ray-$\gamma$-ray correlation supports the idea that emission observed in the two energy regimes is produced by a single electron population and can be described with one zone synchrotron-self-Compton model. 
However, we remind here that PKS 2155-304 is also known from the cubic relation observed between these two energy regimes [7].

Interestingly, the comparisons of optical and VHE $\gamma$-ray fluxes and optical and X-ray fluxes do not show any significant correlation. 
However, we note here that in the case of optical-$\gamma$-ray diagram two trends are visible.
These isolated trends in the diagram correspond to the different outbursts observed in \pks. 
In order to illustrate these different colors are used: orange for the orphan flare, and blue for the rest.

The presence of the orphan flare implies that there is no global explanation for the flux variation. If, for instance, the flux variations were due to changes in particle number at all times, the correlations should show the expected flux-squared behavior between synchrotron and SSC fluxes. The absence of such a global correlation implies that it is not a simple variation in particle number, but a more complex interplay of a larger number of parameters.
On the other hand, the presence of correlations on shorter time scales implies that on these time scales or even for specific flares, simple correlations (and therefore simple parameter variations) hold. But they change from time interval to time interval, so that the correlations on short time scales is
washed out over long time scales.

\begin{figure*}
 \centering{\includegraphics[width=0.245\textwidth]{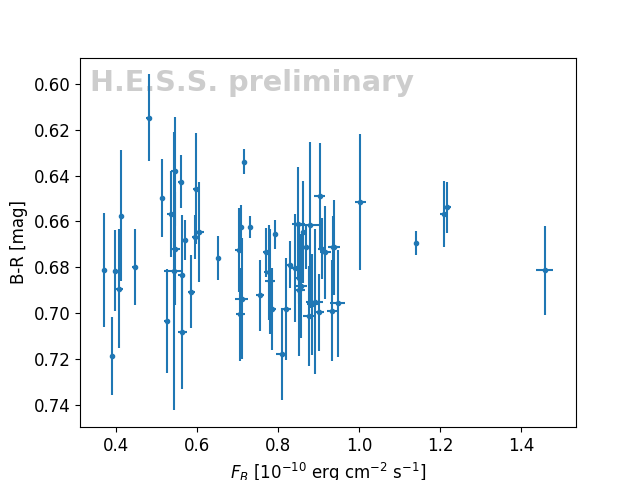}}
 \centering{\includegraphics[width=0.245\textwidth]{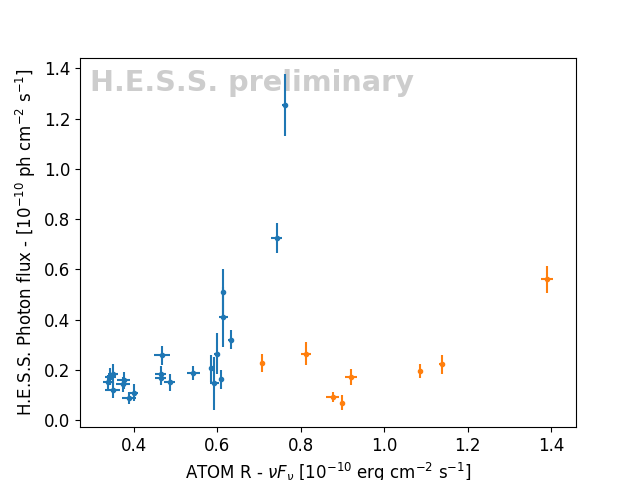}}
 \centering{\includegraphics[width=0.245\textwidth]{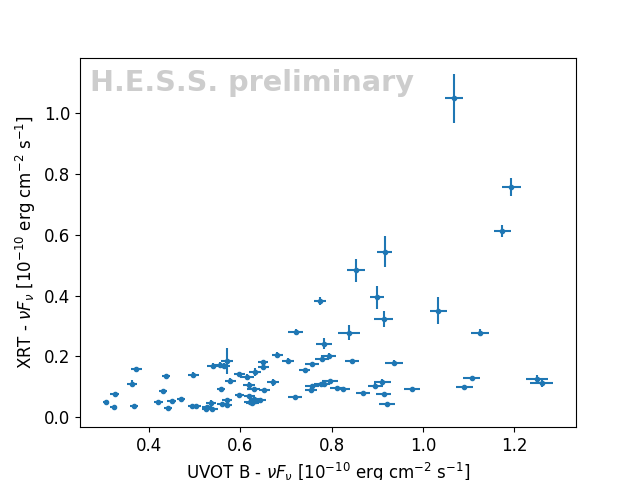}}
 \centering{\includegraphics[width=0.245\textwidth]{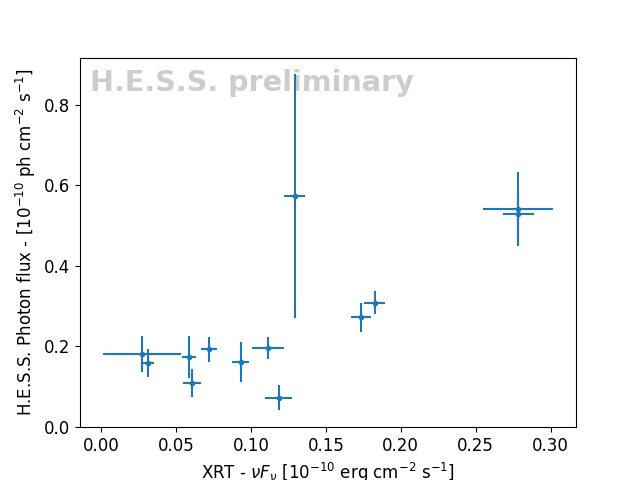}}
  \caption[]{
  Comparisons of emission observed at different wavelengths. From left to right: B-R color vs. the B-band energy flux for optical ATOM observations of \pks; 
  VHE $\gamma$-ray photon flux of the source also as a function of the R-band energy flux. The orphan flare points are denoted with orange points; Comparison of X-ray flux in the energy range of 0.3-10 keV and optical flux in B band as observed with Swift-XRT and Swift-UVOT, respectively; Comparison of X-ray flux as seen with Swift-XRT and VHE $\gamma$-ray flux as seen with H.E.S.S.}
 \label{corr}
\end{figure*}

All comparison plots mentioned above are presented in Figure\,\ref{corr}.

\section{Spectral energy distributions}
Different activity states observed during the monitoring period allow one to perform the study on the spectral variability of the blazar. 
Four different intervals have been selected in order to compare the spectral variability of \pks.
All intervals are marked in the light curve plot (see Figure\,\ref{lc}). 
The selected intervals present: low state in 2013 (hereafter interval 1), post flare observations in 2014 (hereafter interval 2), flaring activity in 2015 (hereafter interval 3), and orphan optical flare in 2016 (hereafter interval 4). 
 For all intervals, the broadband spectral energy distribution was reproduced using a steady-state, one zone code \cite{Bottcher2013}.

Given the input parameters, such as the size of the emission region $R$, the strength of the tangled magnetic field B, the electron injection
luminosity $L_{inj}$, the injection spectral index $s$, and the minimum and maximum electron Lorentz factor $\gamma_{min}$ and $\gamma_{max}$,
respectively, the electron distribution function is self-consistently derived from the competition of injection, cooling, and escape.
In all simulations, the Doppler factor $\delta$ and the escape time parameter $\eta$ (defined as $t_{esc}=\eta
R/c$ with the speed of light $c$) are fixed to $\delta=30$ and $\eta=100$, respectively.
Subsequently, the synchrotron and inverse-Compton spectra are calculated.
In the case of \pks, only synchrotron-self Compton is considered for the inverse-Compton process owing to the absence of strong external photon fields.

\begin{figure*}
 \centering{\includegraphics[width=0.45\textwidth]{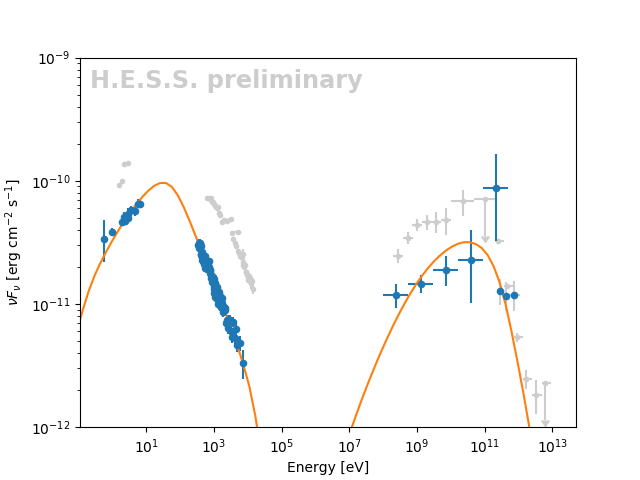}}
 \centering{\includegraphics[width=0.45\textwidth]{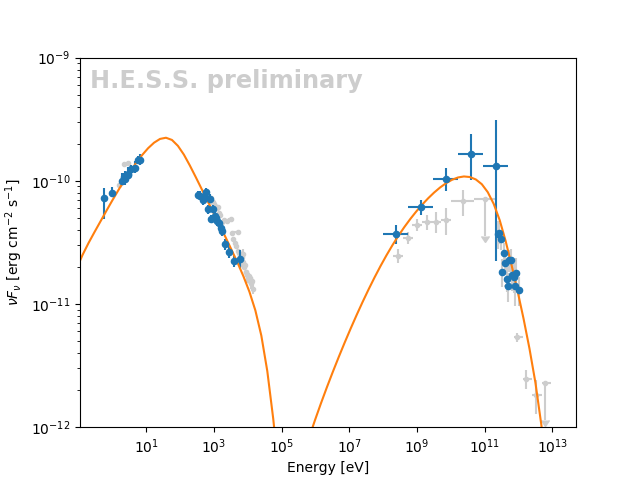}} \\
 \centering{\includegraphics[width=0.45\textwidth]{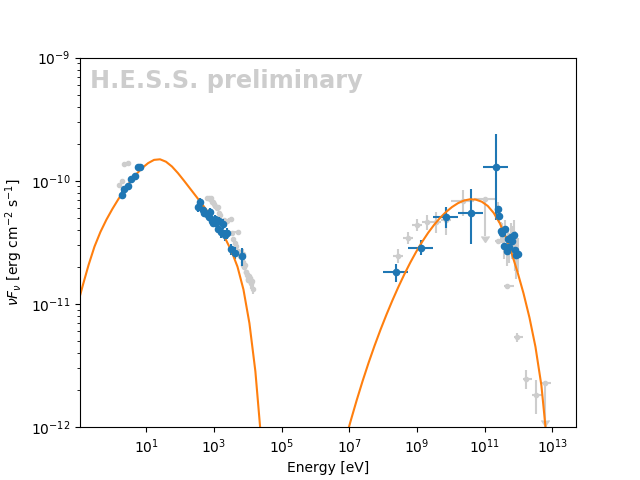}}
 \centering{\includegraphics[width=0.45\textwidth]{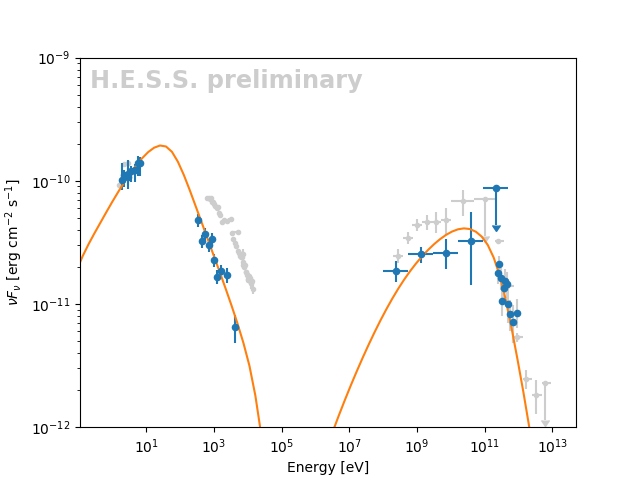}} \\
 \label{SEDs}
  \caption[]{Broadband spectral  energy distributions of \pks\ for four differnt intervals. For top left to bottom right SED plots correspond to interval 1, 2, 3, and 4, respectively.
  In grey the ``2008 low state'' data   are presented \cite{2155_optgam}. }
\end{figure*}

All SEDs  are presented in Figure\,\ref{SEDs}.
As it is presented in Table\,\ref{table:models}.
model parameters differ from one SED to another. 
\begin{itemize}
 \item [--] The Compton dominance, which represents the ratio of high energy peak in the SED to the low energy one, changes from 0.21 (for interval 4) to 0.47 (interval 3). 

The smallest value corresponds to the orphan optical/UV flare. 

\item [--] X-ray spectra follow harder-when-brighter behavior. Such a trend is known for HBL type blazars and was previously reported for the blazar \cite{Zhang2006, Abramowski2012}.

\item [--] SEDs for intervals 1 and 4 are characterized by similar X-ray and $\gamma$-ray spectra. Both have similar B values and $\gamma_{min}$, but doubling of  L$_{inj}$
and R, which leads to a decrease in particle density by factor 4.
As the electron distribution remains soft, most additional particles are added at lower energies, which explains  optical flare

\item [--] SEDs for intervals 2 and 3 are characterized by similar X-ray fluxes, but different X-ray spectral indices. 
Correlated activity in optical and high energy $\gamma$-ray spectra are also observed. 
The modeling parameters differ strongly in B (for interval 2 it is 4 times bigger than for 3) and R (for interval 2 it is 3 times smaller than for 3).
Only minor variations in $\gamma_{min}$ and L$_{inj}$ are present. 

\end{itemize}

The above models can be thought of as a "proof-of-concept", where the steady-state one-zone model is fitted to the data without yet  considering a global physical scenario.

\begin{table}
\begin{tabular}{c|c|c|c|c|c|c}
  \hline
  Interval & L$_{inj}$ & $\gamma_{min}$ & $\gamma_{max}$ & B &  R & s\\
       & 10$^{45}$\,erg$/$s   &  10$^{4}$ & 10$^{6}$ & G  & 10$^{16}$\,cm & \\
  \hline  
  \hline
  interval 1  &  1.6  &  5.0 &  1  & 0.03 &   7.3 & 3.4 \\
  interval 2  &  4.2  &  4.0 &  1  &  0.05 &   5.6 & 3.1\\
  interval 3  & 3.0  & 5.2 &   1   & 0.015 &  1.6 & 2.7  \\
  interval 4  & 3.0  & 4.8 &    1  & 0.03 &   13 & 3.5\\
  \hline
\end{tabular}
\caption{Summary of the modelling parameters. }
\label{table:models}
\end{table}

\section{Summary}
The intensive monitoring of the blazar \pks\ performed in 2013-2016 with H.E.S.S. and accompanying instruments unraveled the very complex nature of the blazar. 
\pks\ is a well-known object thanks to two famous flares observed in 2006, which have not been seen again over the next years of monitoring, however, the behavior of the blazar is still surprising and changes from epoch to epoch. 
The main findings from the monitoring  can be summarized as follows:
\begin{itemize}
\item [--] The rich dataset collected in 2013-2016 is a perfect tool to test MWL correlations in different activity states of \pks.
 These studies have shown the unique behavior of the blazar, including lack of global color-magnitude correlation in the optical data that was previously reported for another set of optical data \cite{2155_atom, Wierzcholska_atom}; lack of global optical-X-ray or $\gamma$-ray-optical correlation, which however is present in shorter sets of data. 
 
\item [--]  \pks\ is the only HBL type blazar for which an orphan optical/UV outburst lasting several months was observed so far. 
We note here that an optical orphan flare was previously reported for the flat spectrum radio quasar  PKS\,0208-512 \cite{Chatterjee_0208}.

\item [--] Due to continuous changes in the behavior of \pks, there seems to be no simple bimodality between a quiescent and a high or flaring state. 
The low state reported for 2008 observation of \pks\ \cite{2155_optgam} correspond to similar flux level in optical or $\gamma$-ray range bands which are now called an outburst.

\end{itemize}

Finally,  it is worth mentioning that such interesting behavior of \pks\ was observed only thanks to dense, regular monitoring of the blazar with H.E.S.S. and accompanying instruments.

\acknowledgments
The full H.E.S.S. acknowledgments can be found at the following link:
https://www.mpi-hd.mpg.de/hfm/HESS/pages/publications/auxiliary/. \\
This paper has made use of up-to-date SMARTS optical/near-infrared light curves that are available at www.astro.yale.edu/smarts/glast/home.php. \\
A.W. is supported by the Polish National Agency for Academic Exchange (NAWA).
M.Z. gratefully acknowledges funding by the German Ministry for Education and Research (BMBF) through grant 05A17PC3.
S.W.  acknowledges support by BMBF through Verbundforschung Astroteilchenphysik, grant number 05A17VH5.

\bibliographystyle{JHEP}
\bibliography{pks}


\end{document}